\documentclass[12pt]{article}
\usepackage{graphicx}
\usepackage[cp1251]{inputenc}
\usepackage{rotating}
\begin{document}
 \noindent {\footnotesize\it Astronomy Letters, 2017, Vol. 43, No 8, pp. 559--566.}
 \newcommand{\dif}{\textrm{d}}

 \noindent
 \begin{tabular}{llllllllllllllllllllllllllllllllllllllllllllll}
 & & & & & & & & & & & & & & & & & & & & & & & & & & & & & & & & & & & & & \\\hline\hline
 \end{tabular}

  \vskip 0.5cm
 \centerline{\bf\large Searching for Stars Closely Encountering with the Solar System}
 \centerline{\bf\large Based on Data from the Gaia DR1 and RAVE5 Catalogues}
 \bigskip
 \bigskip
  \centerline
 {
 V.V. Bobylev\footnote [1]{e-mail: vbobylev@gao.spb.ru} and
 A.T. Bajkova
 }
 \bigskip
 {
 \small\it
 Central (Pulkovo) Astronomical Observatory, Russian Academy of Sciences,

 Pulkovskoe sh. 65, St. Petersburg, 196140 Russia
 }
 \bigskip
 \bigskip
 \bigskip

 {
{\bf Abstract}---We have searched for the stars that either
encountered in the past or will encounter in the future with the
Solar system closer than 2 pc. For this purpose, we took more than
216 000 stars with the measured proper motions and trigonometric
parallaxes from the Gaia DR1 catalogue and their radial velocities
from the RAVE5 catalogue. We have found several stars for which
encounters closer than 1 pc are possible. The star GJ 710, for
which the minimum distance is $d_m=0.063\pm0.044$ pc at time
$t_m=1385\pm52$ thousand years, is the record-holder among them.
Two more stars, TYC 8088-631-1 and TYC 6528-980-1, whose encounter
parameters, however, are estimated with large errors, are of
interest.
  }

\medskip
DOI: 10.1134/S1063773717080011

 \subsection*{INTRODUCTION}
According to the hypothesis of Oort (1950), the Solar system is
surrounded by a comet cloud. Although there are little reliable
data on this cloud, it is highly likely that it has a spherical
shape and a radius $1\times10^5$~AU (0.49 pc). The total number of
comets is supposed to be $10^{11}$. The flybys of Galactic field
stars near the Oort cloud can trigger the formation of comet
showers moving into the region of the major planets (Hills 1981).
In the long run, the possibility that the Moon and the Earth are
bombarded with such comets is not ruled out (Wickramasinghe and
Napier 2008).

The long-term evolution of the Oort cloud was considered on the
basis of numerical simulations, for example, in Emel’yanenko et
al. (2007), Leto et al. (2008), Rickman et al. (2008), and
Dybczy\'nski and Kr\'olikowska (2011). In particular, the
Jupiter--Saturn system was shown to be a tangible barrier leading
to a redistribution of the density of comets in the cloud. Apart
from the flybys of stars, the Oort cloud is perturbed by giant
molecular clouds and the gravitational tide produced by the
Galactic attraction (Dybczy\'nski 2002, 2005; Martinez-Barbosa et
al. 2017).

Matthews (1994), M\"ull\"ari and Orlov (1996), Garcia-S\'anchez et
al. (1999, 2001), Bobylev (2010a, 2010b), Anderson and Francis
(2012), Dybczy\'nski and Berski (2015), Bailer-Jones (2015), Feng
and Bailer-Jones (2015), and Mamajek et al. (2015) searched for
the close encounters of stars with the solar orbit using various
observational data. As a result, $\sim$200 Hipparcos (1997) stars
that either encountered or would encounter with the Solar system
closer than 5 pc in the time interval from $-$10 to +10 Myr were
revealed. Several candidates have a high probability of their
penetration into the Oort cloud region.

For example, the star HIP 85605 (a dwarf of spectral type
$\sim$K4) may encounter with the Solar system within a distance
$d_m\sim0.1$ pc at tm $\sim$330 thousand years (Bailer-Jones
2015); for the star HIP 63721 (F3V) these parameters are
$d_m\sim0.2$ pc and $t_m\sim150$ thousand years (Bailer-Jones
2015; Dybczy\'nski and Berski 2015). At the same time, all authors
point out that the parallaxes of the stars HIP 85605 and HIP 63721
are very unreliable.

The low-mass binary system WISE J072003.20-084651.2 (M9.5 + T5)
with a total mass of $\sim0.15 M_\odot$ is of interest. Mamajek et
al. (2015) estimated $d_m=0.25^{+0.11}_{-0.07}$ pc and
$t_m=-70^{+0.15}_{-0.10}$ thousand years for it.

The star GL 710 (K7V), for which the encounter parameters found
from the Hipparcos data are $d_m=0.31\pm0.17$ pc and
$t_m=1447\pm60$ thousand years (Garcia-S\'anchez et al. 2001;
Bobylev 2010a), is well known. Completely new estimates of these
parameters have recently been obtained by Berski and Dybczy\'nski
(2016) using the parallaxes and proper motions measured in the
Gaia experiment (Prusti et al. 2016): $d_m=0.065\pm0.030$ pc and
$t_m=1350\pm50$ thousand years. Thus, the star GJ 710 remains the
record-holder in terms of encounters among the candidate stars
with more or less reliable measurements.

New possibilities in searching for stars closely encountering with
the Solar system are associated with the appearance of the first
version of the Gaia catalogue. This catalogue was produced from a
combination of the data in the first year of Gaia observations
with the positions and proper motions of Tycho-2 stars (Hog et al.
2000). It is designated as TGAS (Tycho--Gaia Astrometric Solution,
Michalik et al. 2015; Brown et al. 2016; Lindegren et al. 2016)
and contains the parallaxes and proper motions of $\sim$2 million
bright stars. The TGAS version has no stellar radial velocities;
therefore, specialized RAVE type catalogues of radial velocities
should be invoked to calculate the total space velocities of
stars.

The goal of this paper is the search for candidate stars closely
encountering with the Sun based on the present-day data on stars.
For this purpose, we use the Gaia DR1 and RAVE5 catalogues (Kunder
et al. 2017). We construct the orbits of stars using an improved
model Galactic gravitational potential (Bajkova and Bobylev 2016).

 \subsection*{DATA}
The random errors of the parameters included in the Gaia DR1
catalogue are either comparable to or smaller than those given in
the Hipparcos and Tycho-2 catalogues. The mean parallax errors are
$\sim$0.3 mas (milliarcseconds). For most stars of the TGAS
version the mean proper motion error is $\sim$1 mas yr$^{-1}$
(milliarcseconds per year), but for quite a few($\sim$94 000)
stars common to the Hipparcos catalogue this error is smaller by
an order of magnitude, 0.06 mas yr$^{-1}$.

The RAVE (RAdial Velocity Experiment) project (Steinmetz et al.
2006) is devoted to determining the radial velocities of many
faint stars. The observations in the southern hemisphere at the
1.2-m Schmidt telescope of the Anglo-Australian Observatory
started in 2003. Five data releases of this catalogue (DR1--DR5)
have been published since then. The mean radial velocity error is
$\sim$3 km s$^{-1}$. The RAVE DR5 version (Kunder et al. 2017)
contains data on 457 588 stars; there is an overlap with the TGAS
catalogue for about half of these stars.

In this paper we use the data set from Hunt et al. (2016), where
the common stars from the TGAS and RAVE DR5 catalogues were
studied. There are trigonometric parallaxes and proper motions
from the TGAS catalogue and radial velocities from the RAVE DR5
catalogue for 216 201 stars in this list. The stars with a
relative distance error of more than 10\% were excluded when the
sample was produced. Hunt et al. (2016) used both photometric
distance estimates from the RAVE catalogue and trigonometric
parallaxes from the TGAS catalogue. In this paper we calculate all
distances to the stars using their trigonometric parallaxes.

 \subsection*{METHODS}
 \subsubsection*{Model Galactic Gravitational Potential}
The expressions for the potentials are considered in a cylindrical
coordinate system ($R,\psi,z$) with the coordinate origin at the
Galactic center. In a rectangular coordinate system $(x,y,z)$ with
the coordinate origin at the Galactic center the distance to a
star (spherical radius) will be $r^2=x^2+y^2+z^2=R^2+z^2$.

The equations of motion for a test particle in an axisymmetric
gravitational potential $\Phi$ can be derived (see Appendix A to
Irrgang et al. (2013)) from the Lagrangian $\pounds$ of the
system:
\begin{equation}
 \begin{array}{lll}
 \pounds(R,z,\dot{R},\dot{\psi},\dot{z})=
 0.5(\dot{R}^2+(R\dot{\psi})^2+\dot{z}^2)-\Phi(R,z).
 \label{Lagr}
 \end{array}
\end{equation}
Introducing the canonical momenta
\begin{equation}
 \begin{array}{lll}
    p_{R}=\partial\pounds/\partial\dot{R}=\dot{R},\quad
 p_{\psi}=\partial\pounds/\partial\dot{\phi}=R^2\dot{\psi},\quad
    p_{z}=\partial\pounds/\partial\dot{z}=\dot{z},
 \label{moments}
 \end{array}
\end{equation}
we will obtain the Lagrange equations as a system of six
first-order differential equations:
 \begin{equation}
 \begin{array}{llllll}
 \dot{R}=p_R,\\
 \dot{\psi}=p_{\psi}/R^2,\\
 \dot{z}=p_z,\\
 \dot{p_R}=-\partial\Phi(R,z)/\partial R +p_{\psi}^2/R^3,\\
 \dot{p_{\psi}}=0,\\
 \dot{p_z}=-\partial\Phi(R,z)/\partial z.
 \label{eq-motion}
 \end{array}
\end{equation}
The fourth-order Runge–Kutta algorithm was used to integrate
Eqs.~(3).

In this paper we use a three-component model Galactic
gravitational potential:
 \begin{equation}
 \Phi=\Phi_b+\Phi_d+\Phi_h,
 \label{model}
 \end{equation}
where the subscripts denote the bulge, disk, and halo,
respectively.

In accordance with the convention adopted in Allen and Santill\'an
(1991), we express the gravitational potential in units of 100
km$^2$ s$^{-2}$, the distances in kpc, and the masses in units of
the Galactic mass $M_{gal}=2.325\times 10^7 M_\odot$,
corresponding to the gravitational constant $G=1.$

The bulge, $\Phi_b(r)$, and disk, $\Phi_d(r(R,z))$, potentials are
represented by the expressions from Miyamoto and Nagai (1975):
 \begin{equation}
 \renewcommand{\arraystretch}{1.2}
  \Phi_b(r)=-\frac{M_b}{(r^2+b_b^2)^{1/2}},
  \label{bulge}
 \end{equation}
 \begin{equation}
 \Phi_d(R,z)=-\frac{M_d}{\{R^2+[a_d+(z^2+b_d^2)^{1/2}]^2\}^{1/2}},
 \label{disk}
\end{equation}
where $M_b$ and $M_d$ are the masses of the components, $b_b,$
$a_d,$ and $b_d$ are the scale lengths of the components in kpc.

The expression for the halo potential was derived by Irrgang et
al. (2013) based on the expression for the halo mass from Allen
and Martos (1986):
\begin{equation}
 \renewcommand{\arraystretch}{3.2}
  m_h(<r) = \left\{
  \begin{array}{ll}\displaystyle
  \frac{M_h(r/a_h)^{\gamma}}{1+(r/a_h)^{\gamma-1}},
  & \textrm{if }  r\leq\Lambda \\\displaystyle
  \frac{M_h(\Lambda/a_h)^{\gamma}}{1+(\Lambda/a_h)^{\gamma-1}}=\textrm{const},
  & \textrm{if } r>\Lambda  \end{array} \right\},
 \label{m-h-I}
 \end{equation}
It slightly differs from that given in Allen and Santill\'an
(1991) and is
\begin{equation}
 \renewcommand{\arraystretch}{3.2}
  \Phi_h(r) = \left\{
  \begin{array}{ll}\displaystyle
  \frac{M_h}{a_h}\biggl( \frac{1}{(\gamma-1)}\ln \biggl(\frac{1+(r/a_h)^{\gamma-1}}{1+(\Lambda/a_h)^{\gamma-1}}\biggr)-
  \frac{(\Lambda/a_h)^{\gamma-1}}{1+(\Lambda/a_h)^{\gamma-1}}\biggr),
  &\textrm{if }   r\leq \Lambda \\\displaystyle
  -\frac{M_h}{r} \frac{(\Lambda/a_h)^{\gamma}}{1+(\Lambda/a_h)^{\gamma-1}}, &\textrm{if }  r>\Lambda,
  \end{array} \right.
 \label{halo-I}
 \end{equation}
where $M_h$ is the mass, $a_h$ is the scale length, the
Galactocentric distance is $\Lambda=200$ kpc, and the
dimensionless coefficient $\gamma=2.0.$

The model parameters $M_b,$ $M_d,$ $M_h,$ $b_b,$ $a_d,$ $b_d,$ and
$a_h$ were taken from Bajkova and Bobylev (2016), where they were
refined based on a large set of present-day observational data.
Their values are given in Table 1.

The circular rotation velocity of the Galaxy at the adopted
Galactocentric distance of the Sun $R_0=8.3$~kpc is 244 km
s$^{-1}$ (Bajkova and Bobylev 2016). The peculiar velocity
components of the Sun relative to the local standard of rest were
taken to be $(U_\odot,V_\odot,W_\odot)_{LSR}=(10,11,7)$ km
s$^{-1}$ based on the results from Bobylev and Bajkova (2014a) in
agreement with the results from Sch\"onrich et al. (2010). We take
into account the Sun’s height above the Galactic plane
$Z_\odot=16$~pc (Bobylev and Bajkova 2016). We neglect the
star--Sun gravitational interaction.

 {\begin{table}[t]                                  
 \caption[]
 {\small\baselineskip=1.0ex Parameters of the model Galactic potential}
 \label{t:potential}
 \begin{center}\begin{tabular}{|c|r|}\hline
 $M_b$ &  386 M$_{gal}$ \\\hline
 $M_d$ & 3092 M$_{gal}$ \\\hline
 $M_h$ & 452 M$_{gal}$ \\\hline
 $b_b$ & 0.2487 kpc   \\\hline
 $a_d$ & 3.67   kpc   \\\hline
 $b_d$ & 0.3049 kpc   \\\hline
 $a_h$ & 1.52   kpc   \\\hline
 \end{tabular}\end{center}\end{table}}

In the case where the spiral density wave is taken into account
(Lin and Shu 1964; Lin et al. 1969), the following term (Fernandez
et al. 2008) is added to the right-hand side of Eq. (2):
\begin{equation}
 \Phi_{sp} (R,\theta,t)= A\cos[m(\Omega_p t-\theta)+\chi(R)].
 \label{Potent-spir}
\end{equation}
Here,
 $$
 A= \frac{(R_0\Omega_0)^2 f_{r0} \tan i}{m},
 \qquad
 \chi(R)=- \frac{m}{\tan i} \ln\biggl(\frac{R}{R_0}\biggr)+\chi_\odot,
 $$
where $A$ is the amplitude of the spiral potential, $f_{r0}$ is
the ratio between the radial component of the force due to the
spiral arms and that due to the general Galactic field, $\Omega_p$
is the angular velocity of the spiral pattern, $m$ is the number
of spiral arms, $i$ is the pitch angle of the arms ($i<0$ for a
wound pattern), $\chi$ is the radial phase of the spiral wave (the
arm center then corresponds to $\chi=0^\circ$), and $\chi_\odot$
is the radial phase of the Sun in the spiral wave.

The following spiral wave parameters were taken as a first
approximation:
 \begin{equation}
 \begin{array}{lll}
 m=4,\quad
 i=-13^\circ,\quad
 f_{r0}=0.05,\quad
 \chi_\odot=-140^\circ,\quad
 \Omega_p=20~\hbox {km s$^{-1}$ kpc$^{-1}$}
 \label{param-spiral}
 \end{array}
 \end{equation}
We used this set of parameters in Bobylev and Bajkova (2014b),
where a broad overview of the parameter selection problem is
given. If necessary, some of them, in particular, $\chi_\odot$ and
$\Omega_p$, can be varied.

\begin{figure}[t]
{\begin{center}
   \includegraphics[width=0.99\textwidth]{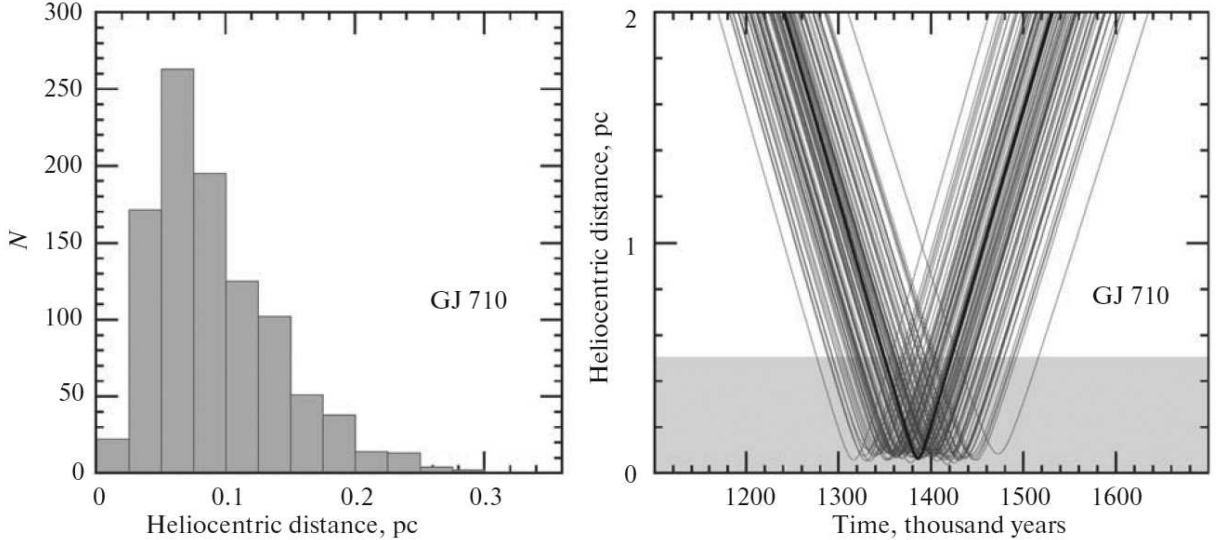}
 \caption{
(a) Histogram of the distribution of model minimum distances $d_m$
obtained for the star GJ 710 by the Monte Carlo method for 1000
trajectories; (b) the thick line indicates the trajectory of GJ
710 relative to the Sun constructed with the parameters from
Table~2 and 100 model trajectories obtained by the Monte Carlo
method, the shading indicates the boundaries of the Oort cloud.
  } \label{f-01}
\end{center}}
\end{figure}
\begin{figure}[t]
{\begin{center}
   \includegraphics[width=0.99\textwidth]{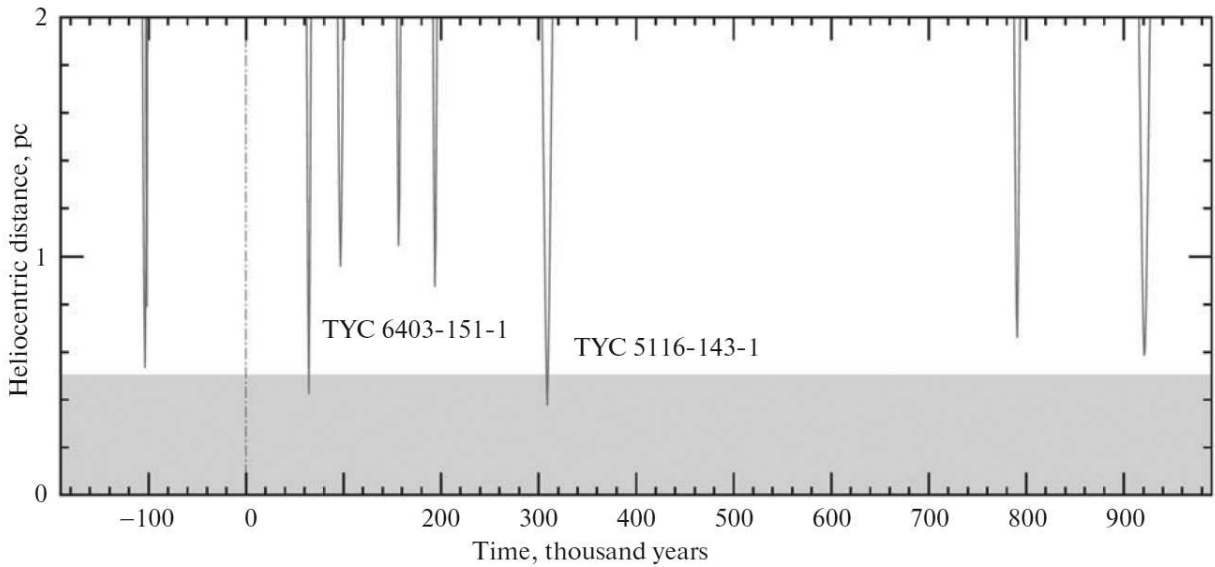}
 \caption{
Trajectories of the nine stars from Table~3, the dash–dotted
vertical line marks the current time, the shading indicates the
boundaries of the Oort cloud.
  } \label{f-02}
\end{center}}
\end{figure}

 \subsubsection*{Monte Carlo Simulations}
In accordance with the method of Monte Carlo simulations, for each
object we calculate a set of orbits by taking into account the
random errors in the input data. For each star we calculate the
parameter of the encounter between the stellar and solar orbits
$d(t)=\sqrt{\Delta x^2(t)+\Delta y^2(t)+\Delta z^2(t)}$. The
closest encounter is characterized by two parameters, $d_m$ and
$t_m.$ The errors of the stellar parameters are assumed to be
distributed as a normal law with a dispersion $\sigma.$ The errors
are added to the stellar equatorial coordinates, proper motion
components, parallaxes, and radial velocities.

 \subsection*{RESULTS AND DISCUSSION}
For each of the 216 201 stars we constructed its orbit relative to
the Sun in the time interval from $-15$ to +15 Myr. From the
entire list we selected the stars with an encounter parameter
$d_m<1$~pc. We divided them into two samples, depending on the
quality of the measured initial velocities and positions.

Sample 1 includes the stars for which the relative errors of the
measured initial velocities and positions do not exceed 10\% and
the measurement error of the radial velocity $<15$~km s$^{-1}$.
The parameters of the stars from sample 1 are given in Table 2,
where columns 1--9 give, respectively, the ordinal star number,
the Tycho identification number (the Hipparcos number is also
provided, if available), the stellar equatorial coordinates
$\alpha$ and $\delta$, the proper motion components
$\mu_\alpha\cos\delta,$ and $\mu_\delta$ with their measurement
errors, the trigonometric parallax with its measurement error, the
radial velocity with its measurement error, the signal-to-noise
(S/N) ratio of the spectrum when determining the radial velocity
copied from column 12 of the RAVE5 catalogue, and the encounter
parameters $d_m$ and $t_m$ we found.

Many stars with huge radial velocities, which is most likely due
to the erroneous measurements in the RAVE catalogue, enter into
the sample. We collected them into separate Table~3. As can be
seen, the S/N ratios for the stars from Table~2 exceed those for
the stars from Table~3 by an order of magnitude.

Several radial velocity determinations are given for some of the
stars in the RAVE5 catalogue. These include, for example, TYC
5116-143-1, TYC 7567-304-1, TYC 5302-849-1, TYC 9163-286-1 or TYC
7978-659-1. The radial velocities usually differ by an order of
magnitude. The encounter parameters calculated for several known
radial velocity measurements are given in Bailer-Jones (2015) in
such cases for each star. In contrast, we give only one value at
which the closest encounter is obtained.

Many of the stars from sample 2 have huge (more than 600 km
s$^{-1}$) space velocities. By this parameter they can be
attributed to hypervelocity stars that are capable of escaping
from the Galactic attractive field. The escape velocity at the
Galactocentric distance of the Sun slightly depends on the model
gravitational potential and is $\sim$550 km s$^{-1}$ (see, e.g.,
Bajkova and Bobylev 2016). The following fact forces us to doubt
that such high velocities are realistic. According to the
well-known Kleiber theorem (Agekyan et al. 1962), the mean
tangential, $V_t,$ and radial, $V_r,$ velocities are related by
the relation $|{\overline V}_t|=0.5\pi|{\overline V}_r|$. Although
this relation is valid in the statistical sense, it does not hold
at all in our case, because for all stars from sample 2
$|V_t|=4.74 r |\mu|<100$ km s$^{-1}$, where
$\mu=\sqrt{\mu^2_\alpha\cos\delta+\mu^2_\delta}.$

Note that the flags $c_1$--$c_{20}$ describing the morphology of
the spectra are specified in the RAVE catalogues. According to
these characteristics, all of the detected stars with radial
velocities $|V_r|>300$ km s$^{-1}$ have very low signal-to-noise
ratios, and the spectra for all these stars are either with
problems in their continuum ($c_{1,2,3}$=``c'') or peculiar
($c_{1,2,3}$=``p''). This leads us to conclude that the radial
velocities of such stars have been measured very poorly.

 \begin{table}[t]                               
 \caption[]{\small\baselineskip=1.0ex\protect
 Data on the stars of sample 1
 }
 \begin{center}\small
 \begin{tabular}{|l|r|r|r|r|c|r|r|r|}\hline
 \label{tab-1}
          TYC &   $\alpha_{J2000}$ & $\mu_\alpha\cos\delta$ & $\pi$  &       $V_r$ & S/N &   $d_m$ & $t_m$     \\
 alternative  &   $\delta_{J2000}$ & $\mu_\delta\quad\quad$ &        &             &     &         & thousand  \\
         name &        deg         &    mas yr$^{-1}$       &   mas  & km s$^{-1}$ &     &     pc  &  years     \\\hline

 5102-100-1   & $274.961836$ & $-0.468\pm0.130$ & $  52.35$ & $ -13.8$ & --- & $   0.063$ & $  1385$ \\
 GJ~710       & $-01.938613$ & $-0.176\pm0.090$ & $\pm0.27$ & $\pm0.3$ &     & $\pm0.044$ & $ \pm52$ \\\hline

 8088-631-1   & $ 87.580551$ & $ 0.353\pm0.670$ & $   4.99$ & $  71.5$ & 64.2 & $   0.37 $ & $-2792$ \\
              & $-45.642797$ & $ 0.037\pm0.940$ & $\pm0.22$ & $\pm1.0$ &      & $\pm1.18 $ & $\pm66$ \\\hline

 6528-980-1   & $107.001787$ & $ 0.221\pm0.605$ & $   4.20$ & $  25.9$ & 25.4 & $   0.86 $ & $-9115$ \\
              & $-25.871894$ & $ 0.625\pm0.599$ & $\pm0.23$ & $\pm4.1$ &      & $\pm5.60 $ & $\pm35$ \\\hline
 \end{tabular}
 \end{center}
 \end{table}
 \begin{table}[t]                               
 \caption[]{\small\baselineskip=1.0ex\protect
 Data on the stars of sample 2 }
 \begin{center}\small\begin{tabular}{|l|r|r|r|r|c|r|r|}\hline
 \label{tab-2}
    TYC      &  $\alpha_{J2000}$ & $\mu_\alpha\cos\delta$ & $\pi$       &    $V_r$ & S/N & $d_m$ & $t_m$   \\
    HIP      &  $\delta_{J2000}$ & $\mu_\delta\quad\quad$ &             &          &     &       &   thousand \\
    (Sp)     &     deg         &    mas yr$^{-1}$          &   mas  & km s$^{-1}$ &     & pc      & years \\\hline

 5116-143-1  &      $278.484540$ &      $ -2.289\pm0.069$ &   $   8.92$ & $  -364$ & 1.2 & $   0.38$ & $  308$ \\
 HIP 91012   &      $ -2.820646$ &      $ -0.005\pm0.052$ &   $\pm0.31$ & $ \pm22$ &     & $\pm0.03$ & $\pm28$ \\
   (F3IV/V)  &&&&&&& \\\hline
 6403-151-1  &      $354.359880$ &      $-14.114\pm1.840$ &   $  18.54$ & $  -851$ & 1.5 & $   0.42$ & $   64$ \\
             &      $-16.826037$ &      $-21.315\pm0.891$ &   $\pm0.56$ & $\pm226$ &     & $\pm0.15$ & $\pm20$ \\\hline
 6622-652-1  &      $153.225519$ &      $-14.047\pm2.788$ &   $  13.14$ & $   741$ & 1.3 & $   0.53$ & $ -103$ \\
             &      $-25.002547$ &      $  3.054\pm0.718$ &   $\pm0.69$ & $ \pm13$ &     & $\pm0.06$ & $ \pm8$ \\\hline
 5508-848-1  &      $170.955510$ &      $  0.312\pm2.297$ &   $   3.24$ & $  -336$ & 1.9 & $   0.59$ & $  920$ \\
             &      $ -8.984829$ &      $ -0.244\pm0.800$ &   $\pm0.74$ & $  \pm8$ &     & $\pm0.75$ & $\pm25$ \\\hline
 8923-1577-1 &      $117.117101$ &      $  0.164\pm0.884$ &   $   2.06$ & $  -614$ & 1.2 & $   0.66$ & $  790$ \\
             &      $-66.890918$ &      $ -0.278\pm0.703$ &   $\pm0.34$ & $ \pm66$ &     & $\pm2.00$ & $ \pm35$ \\\hline
 8822-592-1  &      $336.227642$ &      $ -5.461\pm0.305$ &   $   5.03$ & $  1969$ & 0.7 & $   0.79$ & $ -101$ \\
             &      $-54.123260$ &      $  3.249\pm0.578$ &   $\pm0.25$ & $ \pm30$ &     & $   \pm7$ & $ \pm8$ \\\hline
 9524-1668-1 &      $325.449945$ &      $  3.626\pm0.621$ &   $   5.84$ & $  -887$ & 1.9 & $   0.87$ & $  193$ \\
     (F2V)   &      $-82.958804$ &      $  4.032\pm0.563$ &   $\pm0.38$ & $      $ &     & $\pm0.11$ & $ \pm5$ \\\hline
 7567-304-1  &      $ 48.605192$ &      $-20.624\pm0.613$ &   $  16.67$ & $  -625$ & 1.7 & $   0.96$ & $   96$ \\
             &      $-44.689565$ &      $ 28.482\pm1.114$ &   $\pm0.23$ & $ \pm25$ &     & $\pm0.10$ & $ \pm5$ \\\hline
 9339-404-1  &      $358.159161$ &      $  9.592\pm0.884$ &   $   6.91$ & $  -929$ & 0.9 & $   1.04$ & $  155$ \\
             &      $-69.790350$ &      $  0.293\pm0.633$ &   $\pm0.25$ & $ \pm29$ &     & $\pm0.10$ & $ \pm9$ \\\hline
  \end{tabular}\end{center}\end{table}

Note the star TYC 4888-146-1 (absent in our tables), for which
four radial velocities found from four good ($c_{1,2,3}$=``n'',
the spectrum of a normal star) spectra taken at different epochs
are given in the RAVE5 catalogue. All four values are close to
$V_r=-15$ km s$^{-1}$, and one value ($V_r=1897$ km s$^{-1}$) was
found from a spectrum with problems in its continuum
($c_{1,2,3}$=``c''). All of this reinforces our attitude to the
stars with huge radial velocities from the RAVE5 catalogue as
problem ones.

The measurement error of the radial velocity $\sigma_{V_r}$ is
unknown for several stars. In such cases, we adopted
$\sigma_{V_r}=\pm30$ km s$^{-1}$ for them when estimating the
errors of the encounter parameters $d_m$ and $t_m$ by the Monte
Carlo method.

Figure 1 gives a histogram of the distribution of model minimum
distances $d_m$ obtained for the star GJ 710 by the Monte Carlo
method for 1000 trajectories, 100 model trajectories of GJ 710
relative to the Sun.

Figure 2 gives the trajectories of the encounter of the nine stars
from Table 3 with the solar orbit. All these trajectories resemble
thin vertical lines attributable to large flyby velocities.
Therefore, the impact of these stars on Oort cloud comets can only
be very brief with minor consequences.

Bailer-Jones (2015) analyzed a large sample of stars with the
radial velocities from the RAVE4 catalogue for close encounters.
For example, for the star TYC 5116-143-1 (HIP 91012) (Table 3) he
found $d_m=0.48$~pc and $t_m=302$ thousand years when using the
radial velocity $V_r=-364$ km s$^{-1}$ (as we did). To integrate
the stellar orbits, Bailer-Jones (2015) used a model Galactic
gravitational potential different from ours. In spite of this, we
can conclude that we have results very close to those obtained by
other authors when using the same observational data. As can be
seen from Table 3, interesting encounter parameters, with small
random errors, were obtained for this star. However, the RAVE5
catalogue provides another radial velocity for it,
$V_r=-36.5\pm18.6$ km s$^{-1},$ obtained from a different but,
just as in the former case, poor spectrum. In addition, according
to the measurements by Nordstr\"om et al. (2004),
$V_r=-16.8\pm0.4$ km s$^{-1}$. We found that with such a radial
velocity the encounter of TYC 5116-143-1 (HIP 91012) would not be
close ($d_m>5$~pc).

The star GJ 710 is known quite well. For example, $T_{\rm eff}=
4109$ K and $\log(g)=4.91$ cm s$^{-2}$ for it (Franchini et al.
2014), i.e., this is an orange dwarf with a mass of
$\sim0.6M_\odot$. Since there is no radial velocity for it in the
RAVE catalogues, we took its previously known value from the
catalogue by Gontcharov (2006). Using the new trigonometric
parallax and proper motions from the Gaia DR1 catalogue, we found
the encounter parameters $d_m=0.063\pm0.044$ pc and
$t_m=1385\pm52$ thousand years, which are in excellent agreement
with the estimates obtained by Berski and Dybczy\'nski (2016)
using the same data and a similar technique for calculating the
Galactic stellar orbits, $d_m=0.065\pm0.030$ pc and
$t_m=1350\pm50$ thousand years.

The following parameters are given in the RAVE5 catalogue for the
other stars from Table~2:
 $T_{\rm eff}=5940$~K, $\log(g)=4.08$ cm s$^{-2}$ for TYC 8088-631-1 and
 $T_{\rm eff}=4750$~K, $\log(g)=4.00$0 cm s$^{-2}$ for TYC 6528-980-1.
They show that these stars are dwarfs.

Previously, Dybczy\'nski (2006) and Jime\'enez-Torres et al.
(2011) performed numerical simulations of the evolution of comet
orbits using the penetration of a star like GJ 710 with a mass of
$0.6M_\odot$ and $d_m=0.3$ pc into the Oort cloud as an example
and found a small stream of comets toward the major planets from
the impact of a model star that is difficult to separate from the
stream of comets caused by a Galactic tide. However, using a
closer encounter parameter, $d_m=0.065$~pc, Berski and
Dybczy\'nski (2016) showed that a noticeable stream with a density
of about ten comets per year with a duration of 2--4 Myr could
emerge.

 \subsection*{CONCLUSIONS}
We searched for the stars that encountered or would encounter with
the Solar system closer than 1 pc. For this purpose, we took more
than 216 000 stars with the measured proper motions and
trigonometric parallaxes from the Gaia DR1 catalogue and their
radial velocities from the RAVE5 catalogue. The orbits were
integrated over the time interval from $-$15 to +15 Myr using a
model Galactic gravitational potential that includes an
axisymmetric part (bulge, disk, and halo)with the addition of a
nonaxisymmetric component that allows for the influence of the
Galactic spiral density wave.

We found the stars for which encounters with the Solar system
closer than 1 pc are possible. For the bulk of this list such an
analysis has been made for the first time. We divided all of the
stars found into two samples.

Sample 1 contains the stars with small errors of the input data
and low radial velocities. The star GJ 710, for which the minimum
distance is $d_m=0.063\pm0.044$~pc at time $t_m=1385\pm52$
thousand years, is the record-holder in this sample. This confirms
the estimates that have recently been obtained for GJ 710 from
similar data by Berski and Dybczy\'nski (2016). The first sample
includes two more stars, TYC 8088-631-1 and TYC 6528-980-1, with
$d_m<1$~pc, which, however, are estimated with large errors. For
example, $d_m=0.37\pm1.18$~pc and $t_m=-2792\pm66$ thousand years
were found for TYC 8088-631-1.

The remaining stars enter into sample 2. They are characterized by
unrealistically large space velocities and their random errors.
This is because poor quality spectra were used for these stars in
the RAVE catalogues. Therefore, the results obtained from the
stars of sample 2 are much less trustworthy than the previous
ones.

 \subsubsection*{ACKNOWLEDGMENTS}
We are grateful to the referee for the useful remarks that
contributed to an improvement of the paper. We are thankful to J.
Hunt for the provided data. This work was supported by the Basic
Research Program P--7 of the Presidium of the Russian Academy of
Sciences, the ``Transitional and Explosive Processes in
Astrophysics'' Subprogram.

 \bigskip\medskip{\bf REFERENCES}
{\small

  1. T.A. Agekyan, B.A. Vorontsov-Vel’yaminov, V.G.
Gorbatskii, A.N. Deich, V.A. Krat, O.A. Melnikov, and V.V.
Sobolev, Course of Astrophysics and Stellar Astronomy (Fizmatgiz,
Moscow, 1962) [in Russian].

 2. C. Allen and M.A. Martos, Rev. Mex. Astron. Astrofis. 13, 137 (1986).

 3. C. Allen and A. Santill\'an, Rev. Mex. Astron. Astrofis. 22, 255 (1991).

4. E. Anderson and C. Francis, Astron. Lett. 38, 331 (2012).

5. C.A.L. Bailer-Jones, Astron. Astrophys. 575, 35 (2015).

6. A.T. Bajkova and V.V. Bobylev, Astron. Lett. 42, 567 (2016).

  7. F. Berski and P.A. Dybczy\'nski, Astron. Astrophys. 595, L10 (2016).

8. V.V. Bobylev, Astron. Lett. 36, 220 (2010a).

9. V.V. Bobylev, Astron. Lett. 36, 816 (2010b).

 10. V. Bobylev and A. Bajkova, Mon. Not. R. Astron. Soc. 441, 142 (2014a).

11. V.V. Bobylev and A.T. Bajkova, Astron. Lett. 40, 352 (2014b).

12. V.V. Bobylev and A.T. Bajkova, Astron. Lett. 42, 1 (2016).

13. A.G.A. Brown, A. Vallenari, T. Prusti, J. de Bruijne, F.
Mignard, R. Drimmel, et al. (GAIA Collab.), Astron. Astrophys.
595, 2 (2016).

14. P.A. Dybczy\'nski, Astron. Astrophys. 396, 283 (2002).

15. P.A. Dybczy\'nski, Astron. Astrophys. 441, 783 (2005).

16. P.A. Dybczy\'nski, Astron. Astrophys. 449, 1233 (2006).

17. P.A. Dybczy\'nski and M. Kr\'olikowska, Mon. Not. R. Astron.
Soc. 416, 51 (2011).

18. P.A. Dybczy\'nski and F. Berski, Mon. Not. R. Astron. Soc.
449, 2459 (2015).

19. V.V. Emel’yanenko, D. J. Asher, and M. E. Bailey, Mon. Not. R.
Astron. Soc. 381, 779 (2007).

20. F. Feng and C.A.L. Bailer-Jones, Mon. Not. R. Astron. Soc.
454, 3267 (2015).

21. D. Fernandez, F. Figueras, and J. Torra, Astron. Astrophys.
480, 735 (2008).

22. M. Franchini, C. Morossi, P. di Marcantonio, M.L. Malagnini,
and M. Chavez, Mon. Not. R. Astron. Soc. 442, 220 (2014).

23. J. Garcia-S\'anchez, R.A. Preston, D. L. Jones, P.R. Weissman,
J.-F. Jean-Francois, D.W. Latham, and R.P. Stefanik, Astron. J.
117, 1042 (1999).

24. J. Garcia-S\'anchez, P.R. Weissman, R.A. Preston, D.L. Jones,
J.-F. Lestrade, D.W. Latham, R.P. Stefanik, and J.M. Paredes,
Astron. Astrophys. 379, 634 (2001).

25. G. A. Gontcharov, Astron. Lett. 32, 759 (2006).

26. J. G. Hills, Astron. J. 86, 1730 (1981).

27. The Hipparcos and Tycho Catalogues, ESA SP--1200 (1997).

28. E. Hog, C. Fabricius, V.V. Makarov, U. Bastian, P.
Schwekendiek, A. Wicenec, S. Urban, T. Corbin, and G. Wycoff,
Astron. Astrophys. 355, L27 (2000).

29. J.A.S. Hunt, J. Bovy, and R.G. Carlberg, Astrophys. J. 832,
L25 (2016).

30. A. Irrgang, B. Wilcox, E. Tucker, and L. Schiefelbein, Astron.
Astrophys. 549, 137 (2013).

31. J.J. Jim\'enez-Torres, B. Pichardo, G. Lake, and H. Throop,
Mon. Not. R. Astron. Soc. 418, 1272 (2011).

32. A. Kunder, G. Kordopatis, M. Steinmetz, T. Zwitter, P.
McMillan, L. Casagrande, H. Enke, J. Wojno, et al., Astron. J.
153, 75 (2017).

33. G. Leto, M. Jakubik, T. Paulech, L. Neslu\v{s}an, and P.A.
Dybczy\'nski, Mon. Not. R. Astron. Soc. 391, 1350 (2008).

34. C.C. Lin and F.H. Shu, Astrophys. J. 140, 646 (1964).

 35. C.C. Lin, C. Yuan and F.H. Shu, Astrophys. J. 155, 721 (1969).

36. L. Lindegren, U. Lammers, U. Bastian, J. Hernandez, S.
Klioner, D. Hobbs, A. Bombrun, D. Michalik, et al., Astron.
Astrophys. 595, 4 (2016).

37. E.E. Mamajek, S.A. Barenfeld, V.D. Ivanov, A.Y. Kniazev, P.
V\"ais\"anen, Y. Beletsky, and H.M.J. Boffin, Astrophys. J. Lett.
800, L17 (2015).

38. C. A. Martinez-Barbosa, L. J\'ylkov\'a, S. Portegies Zwart,
and A.G. A. Brown, Mon. Not. R. Astron. Soc. 464, 2290 (2017).

 39. R. A. J. Matthews, R. Astron. Soc. Quart. J. 35, 1 (1994).

 40. D. Michalik, L. Lindegren, and D. Hobbs, Astron. Astrophys. 574, 115 (2015).

 41. M. Miyamoto and R. Nagai, Publ. Astron. Soc. Jpn. 27, 533 (1975).

 42. A. A. M\"ull\"ari and V. V. Orlov, Earth, Moon, Planets 72, 19 (1996).

43. B. Nordstr\"om, M. Mayor, J. Andersen, J. Holmberg, F. Pont,
B. R. Jorgensen, E. H. Olsen, S. Udry, and N. Mowlavi, Astron.
Astrophys. 418, 989 (2004).

44. J.H. Oort, Bull. Astron. Inst. Netherland 11 (408), 91 (1950).

45. T. Prusti, J.H. J. de Bruijne,A.G. A. Brown, A. Vallenari, C.
Babusiaux, C. A. L. Bailer-Jones, U. Bastian, M. Biermann, et al.
(GAIA Collab.), Astron. Astrophys. 595, 1 (2016).

46. H. Rickman, M. Fouchard, C. Froeschl\'e, and G. B. Valsecchi,
Sel. Mech. Dyn. Astron. 102, 111 (2008).

47. R. Sch\"onrich, J. Binney, and W. Dehnen, Mon. Not. R. Astron.
Soc. 403, 1829 (2010).

48. M. Steinmetz, T. Zwitter, A. Siebert, F. G. Watson, K. C.
Freeman, U. Munari, R. Campbell, M. Williams, et al., Astron. J.
132, 1645 (2006).

49. J. T. Wickramasinghe and W. M. Napier, Mon. Not. R. Astron.
Soc. 387, 153 (2008).

}
\end{document}